\documentclass
[aps,amssymb,amsmath,floats,showpacs,twocolumn,pre,superscriptaddress]{revtex4}%
\usepackage{amsfonts}
\usepackage{amsmath}
\usepackage{amssymb}
\usepackage{epsf}
\usepackage{graphicx}
\usepackage{graphics}%
\begin{document}
\title{Organization of modular networks}
\author{S. N. Dorogovtsev}
\email{sdorogov@ua.pt}
\affiliation{Departamento de F{\'{\i}}sica da Universidade de Aveiro, 3810-193 Aveiro, Portugal}
\affiliation{A. F. Ioffe Physico-Technical Institute, 194021 St. Petersburg, Russia}
\author{J. F. F. Mendes}
\email{jfmendes@ua.pt}
\affiliation{Departamento de F{\'{\i}}sica da Universidade de Aveiro, 3810-193 Aveiro, Portugal}
\author{A. N. Samukhin}
\email{samukhin@ua.pt}
\affiliation{Departamento de F{\'{\i}}sica da Universidade de Aveiro, 3810-193 Aveiro, Portugal}
\affiliation{A. F. Ioffe Physico-Technical Institute, 194021 St. Petersburg, Russia}
\author{A. Y. Zyuzin}
\email{A.Zyuzin@mail.ioffe.ru}
\affiliation{A. F. Ioffe Physico-Technical Institute, 194021 St. Petersburg, Russia}

\begin{abstract}
We examine the global organization of heterogeneous equilibrium networks
consisting of a number of well distinguished interconnected
parts---``communities'' or modules. We develop an analytical approach allowing
us to obtain the statistics of connected components and an intervertex
distance distribution in these modular networks, 
and to describe their global organization and structure. In particular, we
study the evolution of the intervertex distance distribution with an
increasing number of interlinks connecting two infinitely large uncorrelated
networks. We demonstrate that even a relatively small number of shortcuts
unite the networks into one. 
In more precise terms, if the number of the interlinks is any finite fraction
of the total number of connections, then the intervertex distance distribution
approaches a delta-function peaked form, and so the network is united.

\end{abstract}

\maketitle

\section{Introduction}

\label{introduction}

\begin{figure}[b]
\par
\begin{center}
\scalebox{0.24}{\includegraphics{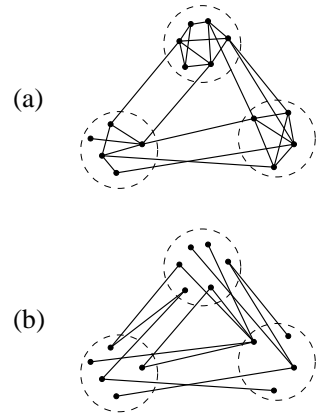}}
\end{center}
\caption{(a) An example of a network, which we study in this paper in the case
$m=3$. The structure of interconnections between the three non-overlapping
subnetworks differs from the structure of connections inside these subnetworks.
Moreover, the structures of the three subnetworks may differ. (b) A contrasting
example of a $3$-partite graph, where connections between vertices of the same
kind are absent. }%
\label{f1}%
\end{figure}

Many real-world networks contain principally distinct parts with different
architectures. In this sense, they are strongly heterogeneous. For example,
the Internet---the net of physically interconnected computers---is connected
to mobile cellular networks. One should note that the issue of the network
heterogeneity is among key problems in the statistical mechanics of complex
networks \cite{ab01a,dm01c,n03,blmch,pvbook04,cbook07,dgm07}. The question is
how do the network's inhomogeneity influence its global structure? The
quantitative description of the global organization of a network is
essentially based on the statistics of $n$-th components of a vertex in the
network, particularly, on the statistics of their sizes \cite{nsw01,n02,dms03}%
. The $n$-th component of a vertex is defined as a set of vertices which are not farther
than distance $n$ from a given vertex. From this statistics, one can easily
find less informative but very useful characteristics---the distribution of
intervertex distances and its first moment, the average intervertex distance.
In the networks with the small-world phenomenon, so-called \textquotedblleft
small worlds\textquotedblright, the mean length of the shortest path
$\overline{\ell}(N)$ between two vertices grows slower than any positive power
of the network size $N$ (the total number of vertices). Rather typically,
$\overline{\ell}(N)\sim\ln N$. As a rule, in infinite small worlds, a
distribution of intervertex distances approaches a delta-function form, where
the mean width $\delta\ell$ is much smaller than $\overline{\ell}$. Moreover,
in uncorrelated networks, $\delta\ell(N\rightarrow\infty)\rightarrow
\text{const}$. So, in simple terms, vertices in these infinite networks are
almost surely mutually equidistant. This statement can be easily understood if
a network has no weakly connected separate parts \cite{remark1}. In this paper
we consider a contrasting situation. Our networks are divided into a number of
non-overlapping but interlinked subnetworks, say $j=1,2,\ldots,m$. What is
important, we suppose that the connections between these subnetworks are
organized differently than inside them, see Fig.~\ref{f1}. This assumption
results in a global (or one may say, macroscopic) heterogeneity of a network.
Using the popular term \textquotedblleft community\textquotedblright, one can
say that our networks have well distinguished communities or modules. Modular
architectures of this kind lead to a variety of effects \cite{ppv06,sh06,ps07,gc07}.
Figure~\ref{f1} explains the difference between these modular networks and the
well studied $m$-partite networks \cite{np03,gl04}.

\begin{figure}[ptb]
\par
\begin{center}
\scalebox{0.26}{\includegraphics{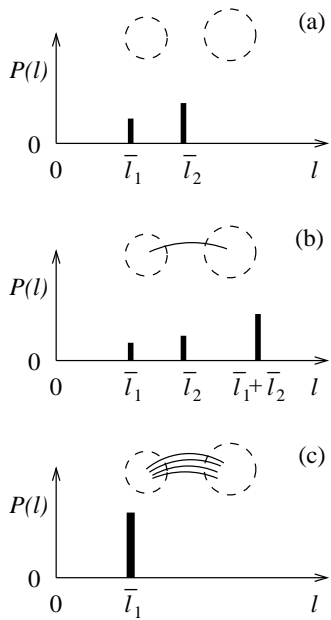}}
\end{center}
\caption{Schematic view of the evolution of an intervertex distance
distribution with the growing number of shortcuts between two large networks:
(a) two separate networks; (b) two networks with a single shortcut between
them; (c) two interconnected networks, when the number of shortcuts is a
finite fraction of the number of edges in the networks. $\overline{\ell}_{1}$
and $\overline{\ell}_{2}$ are the average intervertex distances in the first
and in the second networks, respectively. }%
\label{f2}%
\end{figure}

In this work we analytically describe the statistics of the $n$-th components
in these networks when all $m$ communities are uncorrelated. For the sake of
brevity, here we consider only the case of $m=2$, i.e., of two networks with
shortcuts between them. As an immediate application of this theory we find a
distribution of intervertex distances. We show how the global architecture of
this (large) network evolves with an increasing number of shortcuts, when two
networks merge into one. The question is: when is the mutual equidistance
property realised? How general is this feature? Figure~\ref{f2} schematically
presents our result. The conclusion is that the equidistance is realized when
the number of shortcuts is a finite fraction of the total number of edges in
the network. This finite fraction may be arbitrary small though bigger than
$0$. In this respect, the large network becomes united at arbitrary small
concentrations of shortcuts.

In Sec.~\ref{results} we briefly present our results. Section~\ref{theory}
describes our general approach to these networks based on the Z-transformation
(generating function) technique. In Sec.~\ref{derivations} we explain how to
obtain the intervertex distribution for the infinitely large networks. In
Sec.~\ref{conclusions} we discuss our results. Finally, for the sake of
clarity, in the Appendix we outline the Z-transformation approach in
application to the configuration model of uncorrelated networks.

\section{Main results}

\label{results}

We apply the theory of Sec.~\ref{theory} to the following problem. Two large
uncorrelated networks, of $N_{1}$ and $N_{2}$ vertices, have degree
distributions $\Pi_{1}(q)$ and $\Pi_{2}(q)$ with converging second moments. We
assume that dead ends are absent, i.e., $\Pi_{1}(1)=\Pi_{2}(1)=0$, which
guarantees that finite connected components are not essential in the infinite
network limit (see Ref.~\cite{mr95,mr98}). $\mathcal{L}$ edges interconnect
randomly chosen vertices of net $1$ and randomly chosen vertices of net $2$.
For simplicity, we assume in this problem that $\mathcal{L}$ is much smaller
than the total number of connections in the network. The question is what is
the form of the intervertex distance distribution? In the infinite network
limit, to describe this distribution, it is sufficiently to know three
numbers: an average distance $\overline{\ell}_{1}$ between vertices of
subnetwork $1$, an average distance $\overline{\ell}_{2}$ between vertices of
subnetwork $2$, and an average distance $\overline{d}$ between a vertex from subnetwork $1$
and a vertex from subnetwork $2$. These three numbers give positions of the
three peaks in the distribution. Examining the variations of these three
distances with $\mathcal{L}$ one can find when the equidistance property takes place.

We introduce the following quantities:
\begin{equation}
K_{1,2}\equiv\sum_{q,r}q(q-1)\Pi_{1,2}(q,r)/\bar{q}_{1,2}.\label{e10}%
\end{equation}
Here $\Pi_{1,2}(q,r)$ are given distributions of vertices of intra-degree $q$
and inter-degree $r$ in subnetworks $1,2$ (see Sec.~\ref{theory} for more
detail). 
$\bar{q}_1$ and $\bar{q}_2$ are mean interdegrees of vertices in subnetworks $1$ and $2$, respectively. 
In terms of $K_{1}$ and $K_{2}$, the generalizations of the mean
branching [$\zeta$ in the standard configuration model, see Eqs.~(\ref{190})
in the Appendix] are
\begin{align}
&  \zeta_{1}=K_{1}+\frac{\mathcal{L}^{2}}{N_{1}N_{2}\bar{q}_{1}\bar{q}_{2}%
}\frac{1}{(K_{1}-K_{2})K_{1}^{2}},\nonumber\\[5pt]
&  \zeta_{2}=K_{2}-\frac{\mathcal{L}^{2}}{N_{1}N_{2}\bar{q}_{1}\bar{q}_{2}%
}\frac{1}{(K_{1}-K_{2})K_{2}^{2}}.\label{e20}%
\end{align}
Here we assume that $K_{1}>K_{2}$ and $\zeta_{1}>\zeta_{2}$. We also suppose
that $\zeta_{1},\zeta_{2}<\infty$. If the resulting average distance
$\overline{\ell}_{1}$ between vertices of subnetwork $1$ is smaller than the
corresponding average distance $\overline{\ell}_{2}$ for subnetwork $2$, then
we obtain asymptotically
\begin{align}
\!\!\!\!\!\!\! &  \overline{\ell}_{1}\cong\displaystyle\frac{\ln N_{1}}%
{\ln\zeta_{1}},\label{e30}\\[5pt]
\!\!\!\!\!\!\! &  \overline{\ell}_{2}\cong\overline{d}+\displaystyle\frac
{1}{\ln\zeta_{2}}\ln\left\{  N_{2}\left[  \zeta_{2}^{\overline{d}%
}+C\mathcal{L}\right]  ^{-1}\right\}  \!,\label{e40}\\[5pt]
\!\!\!\!\!\!\! &  \overline{d}\cong\overline{\ell}_{1}+\displaystyle\frac
{\ln(N_{2}/\mathcal{L})}{\ln\zeta_{2}}.\label{e50}%
\end{align}
Here the constant $C$ is determined by the degree distributions $\Pi_{1}(q)$
and $\Pi_{2}(q)$ and is independent of $N_{1}$, $N_{2}$, and $\mathcal{L}$. In
formula~(\ref{e40}), $\zeta_{2}^{\overline{d}}=(N_{1}N_{2}/\mathcal{L}%
)^{\ln\zeta_{2}/\ln\zeta_{1}}$. Note that these asymptotic estimates ignore
constant additives. Formulas~(\ref{e30})--(\ref{e50}) demonstrate that if
$\mathcal{L}$ is a finite fraction of the total number of connections (in the
infinite network limit), then $\overline{\ell}_{2}$ and $\overline{d}$
approach $\overline{\ell}_{1}$. The differences are only finite numbers.
Indeed, the second terms in relations~(\ref{e40}) and (\ref{e50}) are finite
numbers if $N_{2}/\mathcal{L}\rightarrow\text{const}$. [When $\mathcal{L}$ is
a finite fraction of the total number of connections, $\zeta_{2}^{\overline
{d}}\ll\mathcal{L}$ in Eq.~(\ref{e40}).] On the other hand, when $\mathcal{L}$
is formally set to $1$, relation~(\ref{e50}) gives $\overline{d}%
=\overline{\ell}_{1}+\overline{\ell}_{2}$, see Fig.~\ref{f2}(b). Furthermore,
assuming $\mathcal{L}/N_{1,2}\rightarrow0$ as $N_{1,2}\rightarrow\infty$, we
have $\overline{\ell}_{2}\cong\ln N_{2}/\ln\zeta_{2}$ according to
relation~(\ref{e40}), since $\zeta_{2}^{\overline{d}}=(N_{1}N_{2}%
/\mathcal{L})^{\ln\zeta_{2}/\ln\zeta_{1}}\gg\mathcal{L}$.

We also consider a special situation where subnetworks $1$ and $2$ are equal,
so $K_{1}=K_{2}\equiv K$, $\bar{q}_{1}=\bar{q}_{2}$, and $N_{1}=N_{2}\equiv
N$. In this case the mean branching coefficients are
\begin{equation}
\zeta_{1,2}=K\pm\frac{\mathcal{L}}{N\bar{q}_{1}}\frac{1}{K}.\label{e60}%
\end{equation}
With these $\zeta_{1}$ and $\zeta_{2}$, the mean intervertex distances have
the following asymptotics: $\overline{\ell}_{1}=\overline{\ell}_{2}\cong\ln
N/\ln\zeta_{1}$ and $\overline{d}\cong\overline{\ell}_{1}+\ln(N/\mathcal{L}%
)/\ln\zeta_{2}$. Formally setting $\mathcal{L}$ to $1$ we arrive at
$\overline{d}=2\overline{\ell}_{1}=2\overline{\ell}_{2}$. One should stress
that all the listed results indicate a smooth crossover from two separate
networks to a single united one: there is no sharp transition between these
two regimes.

\section{Statistics of modular networks}

\label{theory}

We consider two interlinked undirected networks, one of $N_{1}$, the other of
$N_{2}$ vertices. The adjacency matrix of the joint network, $\hat{g}$, has the following structure:%
\[
\hat{g}=%
\begin{bmatrix}
\hat{g}_{1} & \hat{h}\\
\hat{h}^{T} & \hat{g}_{2}%
\end{bmatrix}
~.
\]
Here $\hat{g}_{1}=\hat{g}_{1}^{T}$ and $\hat{g}_{2}=\hat{g}_{2}^{T}$ are
$N_{1}\times N_{1}$ and $N_{2}\times N_{2}$ adjacency matrices of the first
and of the second subnetworks, respectively, and $\hat{h}$ is $N_{1}\times N_{2}$
matrix for interconnections. We use the following notations: latin (greek)
subscripts $i$, $j$, etc. ($\alpha$, $\beta$, etc.) take values $1,2,\dots,N_{1}$ ($N_{1}+1,N_{1}+2,\dots N_{1}+N_{2}$). So, $g_{ji}$, $g_{\beta\alpha}$,
$g_{j\alpha}$ and $g_{\beta i}$ are the matrix elements of $\hat{g}_{1}$,
$\hat{g}_{2}$, $\hat{h}$ and $\hat{h}^{T}$, resp. We assume the whole network
to be a simple one, i.e., the matrix elements of $\hat{g}$ are either $0$ or
$1$, and the diagonal ones are all zero, $g_{ii}=g_{\alpha\alpha}=0$. 

Every vertex in this network has intra-degree and inter-degree. Vertex $i$
belonging to subnetwork $1$ has intra-degree $q_{i}=\sum_{j}g_{ji}$ and
inter-degree $r_{i}=\sum_{\beta}g_{\beta i}$. Vertex $\alpha$ belonging to
subnetwork $2$ has intra-degree $q_{\alpha}=\sum_{\beta}g_{\beta\alpha}$ and
inter-degree $r_{\alpha}=\sum_{j}g_{j\alpha}$. The total numbers of intra- and
interlinks are $2L_{1}=\sum_{j,i}g_{ij}$, $2L_{2}=\sum_{\beta,\alpha}%
g_{\beta\alpha}$ and $\mathcal{L}=\sum_{j,\alpha}g_{j\alpha}=\sum_{\beta
,i}g_{\beta i}$.

We introduce a natural generalization of the configuration model (we recommend
that a reader look over Appendix to recall the configuration model and the
standard analytical approach to the statistics of its components). In our random network, intralinks in
subnetworks $1$ and $2$ are uncorrelated, and the set of interlinks connecting
them is also uncorrelated. As in the configuration model, our statistical
ensemble includes all possible networks with given sequences of intra- and
inter-degrees for both subnetworks. All the members of the ensemble are taken
with the same statistical weight. Namely, there are $\mathcal{N}_{1,2}\left(
N_{1},N_{2};q,r\right)  $ vertices in subnetworks $1,2$ of intra-degree $q$
and inter-degree $r$. Here $\sum_{q,r}\mathcal{N}_{1,2}\left(  N_{1}%
,N_{2};q,r\right)  =N_{1,2}$. The condition $\sum_{q,r}r\mathcal{N}_{1}\left(
N_{1},N_{2};q,r\right)  =\sum_{q,r}r\mathcal{N}_{2}\left(  N_{1}%
,N_{2};q,r\right)  =\mathcal{L}$, where $\mathcal{L}$ is the number of interlinks, should be
fulfilled. We assume that in the thermodynamic limit $N_{1}\rightarrow\infty$,
$N_{2}\rightarrow\infty$, $N_{2}/N_{1}\rightarrow\kappa<\infty$, we have
$\mathcal{N}_{1,2}\left(  N_{1},N_{2};q,r\right)  /N_{1,2}\rightarrow\Pi
_{1,2}\left(  q,r\right)  $, where $\Pi_{1}$ and $\Pi_{2}$ are given
distribution functions. Again, there is a condition that the number of edges from
subnetwork $1$ to $2$ is the same as from $2$ to $1$:
\begin{equation}
\bar{r}_{1}\equiv\sum_{q,r=0}^{\infty}r\Pi_{1}\left(  q,r\right)  =\kappa
\sum_{q,r=0}^{\infty}r\Pi_{2}\left(  q,r\right)  \equiv\kappa\bar{r}%
_{2}.\label{10}%
\end{equation}
Here $\bar{r}_{1,2}$ are average inter-degrees of the vertices in subnetworks
$1$ and $2$.

The theory of uncorrelated networks extensively uses the Z-representation
(generating function) of a degree distribution:
\begin{equation}
\phi\left(  x\right)  =\sum_{q=0}^{\infty}\Pi\left(  q\right)  x^{q}%
.\label{15}%
\end{equation}
Here we introduce
\begin{equation}
\phi_{1,2}\left(  x,y\right)  =\sum_{q,r=0}^{\infty}\Pi_{1,2}\left(
q,r\right)  x^{q}y^{r}.\label{20}%
\end{equation}
In Z-representation, the average intra- $\bar{q}_{1}$ and $\bar{q}_{2}$ and
inter- $\bar{r}_{1}$ and $\bar{r}_{2}$ degrees of subnetworks $1$ and $2$,
respectively, are
\begin{align}
\!\!\!\!\!\!\!\!\!\!\!\!\!\bar{q}_{1,2}  & {=}\!\left.  \frac{\partial
\phi_{1,2}(x,y)}{\partial x}\right\vert _{x{=}y=1}\!,\ \bar{r}_{1,2}=\,\left.
\frac{\partial\phi_{1,2}(x,y)}{\partial y}\right\vert _{x{=}y=1}\!.\nonumber\\[5pt]
& \label{30}%
\end{align}

Let $\left(  j,i\right)  $, $\left(  \beta,\alpha\right)  $, $\left(
j,\alpha\right)  $ and $\left(  \beta,i\right)$ be ordered vertex pairs. Let
us name their elements in the first and second position as final and initial,
respectively. An \emph{end vertex degree distribution} is the conditional probability
for the 
final vertex of some (randomly chosen) 
ordered pair of vertices   
to have intra- and interdegrees $q$ and $r$, respectively, provided the vertices in this pair are connected by an
edge. We have four distributions, each one depending on two variables:
\begin{align}
P_{1}(q,r)  &  =\frac{1}{2L_{1}}\sum_{ji}\left\langle g_{ji}%
\delta(  q_{j}-1-q)  \delta(  r_{j}%
-r)  \right\rangle ,\nonumber\\[5pt]
P_{2}(q,r)  &  =\frac{1}{2L_{2}}\sum_{\beta,\alpha}\left\langle
g_{\beta\alpha}\delta(  q_{\beta}-1-q)  \delta(
r_{\beta}-r)  \right\rangle ,\nonumber\\[5pt]
Q_{1}(q,r)  &  =\frac{1}{\mathcal{L}}\sum_{j,\alpha}\left\langle
g_{j\alpha}\delta(  q_{j}-q)  \delta(  r_{j}-
1-r)  \right\rangle ,\nonumber\\[5pt]
Q_{2}(q,r)  &  =\frac{1}{\mathcal{L}}\sum_{\beta,i}\left\langle
g_{\beta i}\delta(  q_{\beta}-q)  \delta(  r_{\beta
}-1-r)  \right\rangle . \label{240}%
\end{align}
Taking into account definitions of vertex degrees, we have 
\begin{eqnarray}
&&
P_{1,2}\left(
q,r\right)  =\left(  q+1\right)  \Pi_{1,2}\left(  q+1,r\right)  /\bar{q}_{1,2}
,
\nonumber
\\[5pt]
&&
Q_{1,2}\left(  q,r\right)  =\left(  r+1\right)  \Pi_{1,2}\left(
q,r+1\right)  /\bar{r}_{1,2}
. 
\label{245}
\end{eqnarray}
In Z-representation these distribution functions
take the following forms:%
\begin{align}
\xi_{1}(x,y)  &  =\frac{1}{2L_{1}}\sum_{j}\left\langle q_{j}x^{q_{j}{-}%
1}y^{r_{j}}\right\rangle =\frac{1}{\bar{q}_{1}}\frac{\partial\phi_{1}%
(x,y)}{\partial x},\nonumber\\[5pt]
\xi_{2}(x,y)  &  =\frac{1}{2L_{2}}\sum_{\beta}\left\langle q_{\beta
}x^{q_{\beta}{-}1}y^{r_{\beta}}\right\rangle =\frac{1}{\bar{q}_{2}}%
\frac{\partial\phi_{2}(x,y)}{\partial x},\nonumber\\[5pt]
\eta_{1}(x,y)  &  =\frac{1}{\mathcal{L}}\sum_{j}\left\langle r_{j}x^{q_{i}%
}y^{r_{j}-1}\right\rangle =\frac{1}{\bar{r}_{1}}\frac{\partial\phi_{1}%
(x,y)}{\partial y},\nonumber\\[5pt]
\eta_{2}(x,y)  &  =\frac{1}{\mathcal{L}}\sum_{\beta}\left\langle r_{\beta
}x^{q_{\beta}}y^{r_{\beta}-1}\right\rangle =\frac{1}{\bar{r}_{2}}%
\frac{\partial\phi_{2}(x,y)}{\partial y}. \label{250}%
\end{align}

Let us introduce the $n$-th components of ordered vertex pairs, $C_{n,ji}$,
$C_{n,\beta\alpha}$, $C_{n,j\alpha}$ and $C_{n,\beta i}$. These components are sets,
whose elements are vertices. 
As is natural, the components are 
empty, if the vertices in a pair
are not connected. The first component is either one-element set consisting
of the final vertex, or empty set. For example, $C_{1,\beta i}$ is either
vertex $\beta$ 
or $\varnothing$. The second component, if
nonempty, contains also all the nearest neighbours of the final vertex, except
the initial one, and so on. 
We have four types of the components of an edge: $C_{n,ji}$, $C_{n,\beta\alpha}$, $C_{n,j\alpha}$ and $C_{n,\beta i}$. They are defined
in a recursive way similarly to the standard configuration model (see
Appendix). Each of these four $n$-th components itself consists of two
disjoint sets: one of vertices in subnetwork $1$, the other---in subnetwork $2$. For
example, $C_{n,ji}=C_{n,ji}^{(1)}\cup C_{n,ji}^{(2)}$.

The sizes of the components are $M_{n,ji}^{\left(  1\right)  }=\left\vert
C_{n,ji}^{\left(  1\right)  }\right\vert $, etc. 
Taking into account the
locally tree-like structure of our network gives%
\begin{align}
M_{n,ji}^{\left(  1,2\right)  }  &  =g_{ji}\left[  \binom{1}{0}+\sum_{k\neq
i}M_{n-1,kj}^{\left(  1,2\right)  }+\sum_{\gamma}M_{n-1,\gamma j}^{\left(
1,2\right)  }\right]  ,\nonumber\\[5pt]
M_{n,\beta\alpha}^{\left(  1,2\right)  }  &  =g_{\beta a}\left[  \binom{0}%
{1}+\sum_{k}M_{n-1,k\beta}^{\left(  1,2\right)  }+\sum_{\gamma\neq\alpha
}M_{n-1,\gamma\beta}^{\left(  1,2\right)  }\right]  ,\nonumber\\[5pt]
M_{n,j\alpha}^{\left(  1,2\right)  }  &  =g_{ja}\left[  \binom{1}{0}+\sum
_{k}M_{n-1,kj}^{\left(  1,2\right)  }+\sum_{\gamma\neq\alpha}M_{n-1,\gamma
j}^{\left(  1,2\right)  }x\right]  ,\nonumber\\[5pt]
M_{n,\beta i}^{\left(  1,2\right)  }  &  =g_{\beta i}\left[  \binom{0}{1}%
+\sum_{k\neq i}M_{n-1,k\beta}^{\left(  1,2\right)  }+\sum_{\gamma
}M_{n-1,\gamma\beta}^{\left(  1,2\right)  }\right]  . \label{270}%
\end{align}

The configuration model 
is uncorrelated random network. So 
all the terms on the right-hand side of each of the four equations
(\ref{270}) are independent random variables. Quantities within each of two
sums in these equations are equally distributed. Their statistical properties are also
independent of the degree distribution of the initial vertex of the edge, i.e., 
of $j$ or $\beta$.

The sizes of the connected components of an edge in different networks [e.g.,
$M_{n,jk}^{\left(  1\right)  }$ and $M_{n,jk}^{\left(  2\right)  }$] are, generally, correlated. So we introduce four joint distribution functions of the component sizes in different networks. In Z-representation they are defined as follows: 
\begin{align}
\psi_{n}^{(1)}(x,y)  &  = \frac{1}{2L_{1}}\left\langle \sum_{j,i}%
g_{ji}x^{M_{n,ji}^{\left(  1\right)  }}y^{M_{n,ji}^{\left(  2\right)  }%
}\right\rangle ,\nonumber\\[5pt]
\psi_{n}^{(2)}(x,y)  &  =\frac{1}{2L_{2}}\left\langle \sum_{\beta,\alpha
}g_{\beta\alpha}x^{M_{n,\beta\alpha}^{(1)}}y^{M_{n,\beta\alpha}^{(2)}%
}\!\right\rangle ,\nonumber\\[5pt]
\theta_{n}^{(1)}(x,y)  &  =\frac{1}{\mathcal{L}}\left\langle \sum_{j,\alpha
}g_{j\alpha}x^{M_{n,j\alpha}^{\left(  1\right)  }}y^{M_{n,j\alpha}^{\left(
2\right)  }}\right\rangle ,\nonumber\\[5pt]
\theta_{n}^{(2)}(x,y)  &  =\frac{1}{\mathcal{L}}\left\langle \sum_{\beta
,i}g_{\beta i}x^{M_{n,\beta i}^{\left(  1\right)  }}y^{M_{n,\beta i}^{\left(
2\right)  }}\right\rangle . \label{280}%
\end{align}

The recursive relations for these distributions are straightforward
generalization of a relation for a usual uncorrelated network, without modularity [see Eq.~(\ref{120}) in the Appendix]%
\begin{align}
\psi_{n}^{\left(  1\right)  }\left(  x,y\right)   &  =x\xi_{1}\left[
\psi_{n-1}^{\left(  1\right)  }\left(  x,y\right)  ,\theta_{n-1}^{\left(
2\right)  }\left(  x,y\right)  \right]  ,\nonumber\\[5pt]
\psi_{n}^{\left(  2\right)  }\left(  x,y\right)   &  =y\xi_{2}\left[
\psi_{n-1}^{\left(  2\right)  }\left(  x,y\right)  ,\theta_{n-1}^{\left(
1\right)  }\left(  x,y\right)  \right]  ,\nonumber\\[5pt]
\theta_{n}^{\left(  1\right)  }\left(  x,y\right)   &  =x\eta_{1}\left[
\psi_{n-1}^{\left(  1\right)  }\left(  x,y\right)  ,\theta_{n-1}^{\left(
2\right)  }\left(  x,y\right)  \right]  ,\nonumber\\[5pt]
\theta_{n}^{\left(  2\right)  }\left(  x,y\right)   &  =y\eta_{2}\left[
\psi_{n-1}^{\left(  2\right)  }\left(  x,y\right)  ,\theta_{n-1}^{\left(
1\right)  }\left(  x,y\right)  \right]  . \label{290}%
\end{align}

The $n$-th component $\mathcal{C}_{n,i}$ ($\mathcal{C}_{n,\alpha}$) of vertex
$i$ ($\alpha$) contains all vertices at distance $n$ from vertex $i$ ($\alpha
$) or closer. Let $\mathcal{M}_{n,i}^{\left(  1,2\right)  }=\left\vert
\mathcal{C}_{n,i}^{\left(  1,2\right)  }\right\vert $ and $\mathcal{M}%
_{n,\alpha}^{\left(  1,2\right)  }=\left\vert \mathcal{C}_{n,\alpha}^{\left(
1,2\right)  }\right\vert $ be sizes of the components [$\mathcal{C}^{\left(
1\right)  }$ and $\mathcal{C}^{\left(  2\right)  }$ are the subset of
$\mathcal{C}$, containing vertices of the first and second networks,
respectively]. Using the locally tree-like structure of the network and
absence of correlations between its vertices, we obtain the Z-transform of the
joint distributions of component sizes:
\begin{align}
\Psi_{n}^{(1)}(x,y)  &  =\frac{1}{N_{1}}\left\langle \sum_{i}x^{\mathcal{M}%
_{n,i}^{(1)}}y^{\mathcal{M}_{n,i}^{(2)}}\right\rangle \nonumber\\[5pt]
&  =x\phi_{1}\left[  \psi_{n}^{\left(  1\right)  }\left(  x,y\right)
,\theta_{n}^{\left(  1\right)  }\left(  x,y\right)  \right]  ,\nonumber\\[5pt]
\Psi_{n}^{(2)}(x,y)  &  =\frac{1}{N_{2}}\left\langle \sum_{\alpha
}x^{\mathcal{M}_{n,\alpha}^{(1)}}y^{\mathcal{M}_{n,\alpha}^{(2)}}\right\rangle
\nonumber\\[5pt]
&  =y\phi_{2}\left[  \psi_{n}^{\left(  2\right)  }\left(  x,y\right)
,\theta_{n}^{\left(  2\right)  }\left(  x,y\right)  \right]  . \label{310}%
\end{align}

The conditional average sizes of the components are expressed in terms of the
derivatives of the corresponding distribution functions at the point $x=y=1$.
For example, the conditional average component sizes for an internal vertex
pair in network $1$ are expressed as follows: for the part, which belongs to
the first network it is
\begin{equation}
\overline{M}_{n}^{\left(  111\right)  }=\left\langle g_{ij}M_{n,ij}^{\left(
1\right)  }\right\rangle /\left\langle g_{ij}\right\rangle =\left.
\frac{\partial\psi_{n}^{\left(  1\right)  }\left(  x,y\right)  }{\partial
x}\right\vert _{x=y=1}%
\label{312}
\end{equation}
for the component part in network $1$, and
\begin{equation}
\overline{M}_{n}^{\left(  211\right)  }=\left\langle g_{ij}M_{n,ij}^{\left(
2\right)  }\right\rangle /\left\langle g_{ij}\right\rangle =\left.
\frac{\partial\psi_{n}^{\left(  1\right)  }\left(  x,y\right)  }{\partial
y}\right\vert _{x=y=1}%
\label{314}
\end{equation}
for the part of the component, which is in network $2$. Here, (i) the first
superscript index indicates whether the component is in subnetwork $1$ or $2$,
(ii) the second superscript index indicates whether the final vertex is in
subnetwork $1$ or $2$, and (iii) the third superscript index indicates whether
the initial vertex is in subnetwork $1$ or $2$. For the components of a pair
with initial vertex in network $2$ and final in network $1$ we have:
\begin{equation}
\overline{M}_{n}^{\left(  112\right)  }=\left\langle g_{i\alpha}M_{n,i\alpha
}^{\left(  1\right)  }\right\rangle /\left\langle g_{i\alpha}\right\rangle
=\left.  \frac{\partial\theta_{n}^{\left(  1\right)  }\left(  x,y\right)
}{\partial x}\right\vert _{x=y=1}
\label{316}
\end{equation}
and%
\begin{equation}
\overline{M}_{n}^{\left(  212\right)  }=\left\langle g_{i\alpha}M_{n,i\alpha
}^{\left(  2\right)  }\right\rangle /\left\langle g_{i\alpha}\right\rangle
=\left.  \frac{\partial\theta_{n}^{\left(  1\right)  }\left(  x,y\right)
}{\partial y}\right\vert _{x=y=1},
\label{318}
\end{equation}
and so on.

Using Eqs.~(\ref{290}) one can derive recurrent relations for the average
values of $M_{n}$ and $M_{n-1}$. We introduce a pair of four-dimensional
vectors:%
\begin{equation}
\mathbf{M}_{n}^{\left(  1\right)  }=%
\begin{pmatrix}
\overline{M}_{n}^{\left(  111\right)  }\\
\overline{M}_{n}^{\left(  112\right)  }\\
\overline{M}_{n}^{\left(  121\right)  }\\
\overline{M}_{n}^{\left(  122\right)  }%
\end{pmatrix}
,\ \ \ \ \mathbf{M}_{n}^{\left(  2\right)  }=%
\begin{pmatrix}
\overline{M}_{n}^{\left(  211\right)  }\\
\overline{M}_{n}^{\left(  212\right)  }\\
\overline{M}_{n}^{\left(  221\right)  }\\
\overline{M}_{n}^{\left(  222\right)  }%
\end{pmatrix}
~.\label{320}%
\end{equation}
Then the recurrent relations take the forms:
\begin{equation}
\mathbf{M}_{n}^{\left(  1\right)  }=\widehat{\boldsymbol{\zeta}}%
\mathbf{M}_{n-1}^{\left(  1\right)  }+\mathbf{m}_{1},\ \ \mathbf{M}_{n}^{\left(
2\right)  }=\widehat{\mathbf{\zeta}}\mathbf{M}_{n-1}^{\left(  2\right)
}+\mathbf{m}_{2}~,\label{330}%
\end{equation}
where
\begin{equation}
\widehat{\boldsymbol{\zeta}}{=}\!%
\begin{pmatrix}
\xi_{11} & 0 & \xi_{21} & 0\\
\eta_{11} & 0 & \eta_{21} & 0\\
0 & \eta_{22} & 0 & \eta_{12}\\
0 & \xi_{22} & 0 & \xi_{12}%
\end{pmatrix}\!\!,\ \ \mathbf{m}_{1}{=}\!%
\begin{pmatrix}
1\\
1\\
0\\
0
\end{pmatrix}
\!\!,\ \ \mathbf{m}_{2}{=}\!\!%
\begin{pmatrix}
0\\
0\\
1\\
1
\end{pmatrix}
\!\!.\label{340}%
\end{equation}
Here%
\[
\xi_{\mu\nu}=\left.  \partial_{\mu}\xi_{\nu}\left(  x,y\right)  \right\vert
_{x=y=1},\ \eta_{\mu\nu}=\left.  \partial_{\mu}\eta_{\nu}\left(  x,y\right)
\right\vert _{x=y=1}, 
\]
$\mu,\nu=1,2$. 
The initial conditions are $\mathbf{M}_{1}^{\left(  1\right)  }=\mathbf{m}%
_{1}$, $\mathbf{M}_{1}^{\left(  2\right)  }=\mathbf{m}_{2}$. As for the
average sizes of the $n$-th components of vertices, they are 
\begin{align}
\overline{\mathcal{M}}_{n}^{(11)} &  =\frac{1}{N_{1}}\left\langle \sum
_{i}\mathcal{M}_{n,i}^{\left(  1\right)  }\right\rangle =\left.
\frac{\partial\Psi_{n}^{(1)}\left(  x,y\right)  }{\partial x}\right\vert
_{x=y=1}\nonumber\\[5pt]
&  =1+\bar{q}_{1}\overline{M}_{n}^{(111)}+\bar{r}_{1}\overline{M}_{n}%
^{(112)},\nonumber\\[5pt]
\overline{\mathcal{M}}_{n}^{(21)} &  =\frac{1}{N_{1}}\left\langle \sum
_{i}\mathcal{M}_{n,i}^{\left(  2\right)  }\right\rangle =\left.
\frac{\partial\Psi_{n}^{(1)}\left(  x,y\right)  }{\partial y}\right\vert
_{x=y=1}\nonumber\\[5pt]
&  =\bar{q}_{1}\overline{M}_{n}^{(211)}+\bar{r}_{1}\overline{M}_{n}%
^{(212)},\nonumber\\[5pt]
\overline{\mathcal{M}}_{n}^{(12)} &  =\frac{1}{N_{1}}\left\langle \sum
_{\alpha}\mathcal{M}_{n,i}^{\left(  1\right)  }\right\rangle =\left.
\frac{\partial\Psi_{n}^{(2)}\left(  x,y\right)  }{\partial x}\right\vert
_{x=y=1}\nonumber\\[5pt]
&  =\bar{q}_{2}\overline{M}_{n}^{(122)}+\bar{r}_{2}\overline{M}_{n}%
^{(121)},\nonumber\\[5pt]
\overline{\mathcal{M}}_{n}^{(22)} &  =\frac{1}{N_{2}}\left\langle \sum
_{i}\mathcal{M}_{n,i}^{(2)}\right\rangle =\left.  \frac{\partial\Psi_{n}%
^{(2)}\left(  x,y\right)  }{\partial y}\right\vert _{x=y=1}\nonumber\\[5pt]
&  =1+\bar{q}_{2}\overline{M}_{n}^{(222)}+\bar{r}_{2}\overline{M}_{n}%
^{(221)}.\label{350}%
\end{align}
Here, (i) the first superscript index of $\overline{\mathcal{M}}_{n}$
indicates whether the component is in subnetwork $1$ or $2$, and (ii) the
second superscript index indicates whether a mother vertex is in subnetwork
$1$ or $2$. Recall that $\bar{q}_{1}$ and $\bar{q}_{2}$ are the mean numbers
of internal connections of vertices in subnetworks $1$ and $2$, respectively;
$\bar{r}_{1}$ is a mean number of connections of a vertex in subnetwork $1$,
which go to subnetwork $2$; and finally $\bar{r}_{2}$ is a mean number of
connections of a vertex in subnetwork $2$, which go to subnetwork $1$.
Relations~(\ref{330}) and (\ref{350}) allow us to obtain the average sizes of
all components. 

{\em Example}.---Since formulas in this section are rather cumbersome, to help the readers, we present a simple demonstrative example of the application of these relations. Let us describe the emergence of a giant connected components in a symmetric situation, where both subnetworks have equal sizes and identical degree distributions $\Pi_{1,2}(q,r)\equiv \Pi(q,r)$. In this case, $\xi_1(x,y)=\xi_2(x,y)\equiv\xi(x,y)$ and $\eta_1(x,y)=\eta_2(x,y)\equiv\eta(x,y)$. Also, $\psi^{(1)}(x,y)=\psi^{(2)}(x,y)\equiv\psi(x,y)$ and $\theta^{(1)}(x,y)=\theta^{(2)}(x,y)\equiv\theta(x,y)$. So the relative size $S$ of a giant connected component takes the form: 
\begin{equation}
S=1-\phi(t,u),
\label{352}
\end{equation}
where $t\equiv\psi(1,1)$ and $u\equiv\theta(1,1)$ are non-trivial solutions of the equations:  
\begin{equation}
t = \xi(t,u), \ \ \ u = \eta(t,u).
\label{354}
\end{equation}

For example, let the subnetworks be classical random graphs, and each vertex has no interlinks with a probability $1-p$ and has a single interlink with the complimentary probability $p$. That is, 
\begin{equation}
\Pi(q,r) = e^{-\bar{q}}\frac{\bar{q}^q}{q!}[(1-p)\delta_{q,0} + p\delta_{q,1}], 
\label{356}
\end{equation}
where $\bar{q}$ is the mean vertex intra-degree, so $\phi(x,y) = e^{\bar{q}(x-1)}[1-p+py]$.  

For a single classical random graph with vertices of average degree $\bar{q}$, Eqs.~(\ref{210}) and (\ref{220}) give the point of the birth of a giant connected component, $\bar{q}=q_c=1$, and the relative size of this component $S \cong 2(\bar{q}-1)$ in the critical region. 

Let us now find the birth point $\bar{q}=q_c(p)$ and the critical dependence $S(\bar{q},p)$ in the modular network. For this network, we find $\xi(x,y)=\partial_x \phi(x,y)/\bar{q}=\phi(x,y)$ and $\eta(x,y)=\partial_y \phi(x,y)/\bar{r}=e^{\bar{q}(x-1)}$. Substituting these functions into Eqs.~(\ref{352}) and (\ref{354}) directly leads to the result: 
\begin{equation}
q_c=\frac{1}{1+p}, \ \ \ S \cong 2\frac{(1+p)^3}{1+3p}\left(\bar{q}-\frac{1}{1+p}\right), 
\label{358}
\end{equation} 
compare with a single classical random graph.

\section{Intervertex distance distribution}

\label{derivations}

As was explained in Sec.~\ref{results}, the intervertex distance distribution
in the thermodynamic limit is completely determined by the three mean
intervertex distances: $\overline{\ell}_{1}$ for subnetwork $1$,
$\overline{\ell}_{2}$ for subnetwork $2$, and $\overline{d}$ for pairs of
vertices where the first vertex is in subnetwork $1$ and the second is in
subnetwork $2$. The idea of the computation of these intervertex distances is
very similar to that in the standard configuration model, see the Appendix,
Eq.~(\ref{230}). However, the straightforward calculations for two
interconnected networks are cumbersome, so here we only indicate some points
in our derivations without going into technical details.

The calculations are based on the solution of recursive relations~(\ref{330}).
As is usual, these relations should be investigated in the range $1-x\ll1$,
$1-y\ll1$ of the Z-transformation parameters. Fortunately, the problem can be
essentially reduced to the calculation of two highest eigenvalues of a single
$4\times4$ matrix. The resulting eigenvalues $\zeta_{1}$ and $\zeta_{2}$ for
networks with $\mathcal{L}/N_{1,2}\ll1$ are given by formulas~(\ref{e20}) and
(\ref{e60}). The $n$-th component sizes are expressed in terms of these
eigenvalues. The leading contributions to the $n$-th component sizes turn out
to be linear combinations of powers of the mean branchings: $A\zeta_{1}%
^{n}+B\zeta_{2}^{n}$. The factors $A$ and $B$ do not depend on $n$. For
example, when $\zeta_{1}>\zeta_{2}$, the main contributions to $\overline
{\mathcal{M}}_{n}^{(11)}$ and $\overline{\mathcal{M}}_{n}^{(21)}$ look as
$\zeta_{1}^{n}+[\mathcal{L}^{2}/(N_{1}N_{2})]\zeta_{2}^{n}$ and $(\mathcal{L}%
/N_{1})\zeta_{2}^{n}$, respectively. Here we omitted non-essential factors and
assumed a large $n$. This approximation is based on the tree ansatz, that is
on the locally tree-like structure of the network. This ansatz works when the
$n$-th components are much smaller than subnetworks $1$ and $2$. So the
intervertex distances are obtained by comparing the sizes of relevant $n$-th
components of vertices with $N_{1}$ and $N_{2}$. Since networks $1$ and $2$
are uncorrelated, this estimate gives only a constant additive error which is
much smaller than the main contribution of the order of $\ln N_{1,2}$. (See
Ref.~\cite{dms03} for complicated calculations beyond the tree ansatz in the
standard configuration model, which allow one to obtain this constant number.) 

One should emphasize an additional difficulty specific for the networks under
consideration. The problem is that in some range of $N_{1}$ and $N_{2}$, and
$\zeta_{1}$ and $\zeta_{2}$, while an $n$-th component in, say, network $1$ is
already of size $\sim N_{1}$ (failing tree ansatz), the corresponding $n$-th
component in network $2$ is still much smaller than $N_{2}$. In terms of
Sec.~\ref{theory}, this, e.g., means that there exists a range of $n$,
$\overline{\ell}_{1}{<}n{<}\overline{d}$, where $\overline{\mathcal{M}}%
_{n}^{(11)}{\sim}N_{1}$ but still $\overline{\mathcal{M}}_{n}^{(21)}{\ll}%
N_{2}$. Computing $\overline{\mathcal{M}}_{n}^{(21)}$ in this regime, we use
the tree ansatz, while $\overline{\mathcal{M}}_{n}^{(11)}$ is set to $N_{1}$.
This approximation also produces only a constant additive error which one may
ignore in these asymptotic estimates.

\section{Discussion and conclusions}

\label{conclusions}

A few points should be stressed.

(i) In Section~\ref{theory} we derived relations for the Z-transformation of
the distributions of $n$-th components. Quite similarly to the standard
configuration model (see Appendix), using these formulas with $n\rightarrow
\infty$ readily gives corresponding relations for the statistics of finite
connected components and for the size of a giant connected component. Note
that when subnetworks $1$ and $2$ are uncorrelated, which is our case, 
finite connected components are essential in an infinite network only if there
is a finite fraction of vertices of degree $1$. 

(ii) The theory of Sec.~\ref{theory} is essentially based on the locally
tree-like structure of networks under consideration. In principle one can go
even beyond the tree ansatz as was done for the standard configuration model
in Ref.~\cite{dms03}. This is a challenging problem for these networks. Since
we extensively used the tree approximation in Sec.~\ref{derivations}, our
results for the intervertex distances are only asymptotic estimates.

(iii) For the sake of brevity, we obtained relations only for networks with
two interlinked subnetworks, but it is not a restriction. A generalization to
networks with an arbitrary number of interlinked subnetworks is
straightforward. The final relations in Sec.~\ref{theory} can be readily
generalized without derivation. Generalization to structured networks with
degree--degree correlations is also clear. Note that, in particular, our theory can describe multi-partite networks, whose subnetworks have no intraconnections. Based on equations derived for the configuration model \cite{fr04,dgm06} (for $k$-cores in real-world networks, see Refs.~\cite{adb06,chk07}), one can also generalize this theory to describe the $k$-core organization of modular networks 

(iv) As an application, we considered interlinked networks with a relatively
small number of shortcuts. Note however that our general results in
Sec.~\ref{theory} do not assume this restriction.

In summary, we have developed an analytical approach to the statistics of
networks with well distinguished communities. We have derived general
relations allowing one to find the distributions of the sizes of connected
components in these networks. As a particular application of this theory, we
have obtained asymptotic estimates for the distribution of intervertex
distances in two weakly interconnected uncorrelated networks. We have shown
that in the infinite network limit, vertices in this network are almost surely
equidistant if the relative number of interlinks is any finite number. Our
approach can be applied to a number of other problems for networks of this
sort, including the birth of a giant connected component, percolation, and others.

\begin{acknowledgments}
This work was partially supported by projects POCTI: FAT/46241/2002,
MAT/46176/2002, FIS/61665/2004, and BIA-BCM/62662/2004, 
and also by project DYSONET-NEST/012911 and the ICT project SOCIALNETS, grant 217141. Authors thank A.~V.
Goltsev and A.~M. Povolotsky for useful discussions.
\end{acknowledgments}

\appendix*

\section{Statistics of the configuration model}

\label{appendix}

For the sake of clarity, here we outline the Z-transformation (generating
function) technique in application to the standard configuration model of an
uncorrelated graph with a given degree distribution $\Pi(q)$. For more detail, see Refs.~\cite{nsw01,n02,dms03}. In
simple terms, the configuration model \cite{b80,bc78} is a maximally random
graph with a given degree distribution. In graph theory it is also called a
random graph with a given degree sequence.

Graph of size $N$ consists of a set of vertices $v_{i}$, $i=1,2,\dots,N$,
connected by edges $e_{ji}$. An edge $e_{ji}$ exists if the adjacency matrix
element $g_{ji}=1$. We start from the following distribution:
\begin{equation}
\widetilde{\Pi}
(q)=\frac{1}{2L}\left\langle \sum_{ji=1}^{N}g_{ji}\delta_{K}\left(
q_{j}-1-q\right)  \right\rangle ,\label{40}%
\end{equation}
where $\delta_{K}$ is the Kronecker symbol. This is the probability that a
randomly chosen end of a randomly chosen edge in the graph has branching $q$.
Alternatively, it may be considered as conditional probability for final
vertex $j$ in a randomly chosen ordered pair $\left(  j,i\right)  $ to have
degree $q+1$, provided vertices are connected by an edge. Obviously,
\begin{equation}
\widetilde{\Pi}
(q)  =\frac{1}{2L}\left\langle \sum
_{j=1}^{N}q_{j}\delta_{K}\left(  q_{j}-1-q\right)  \right\rangle =\frac
{q+1}{\bar{q}}\Pi(q+1)  ,\label{50}%
\end{equation}
where $\bar{q}\equiv\langle q\rangle$ is the average degree of a vertex. In the
Z-representation this relation takes the form:%
\begin{equation}
\widetilde{\phi}
(x)  =\sum_{q=0}^{\infty}
\widetilde{\Pi}
(q)  x^{q}=\frac{1}{2L}\left\langle \sum_{j=1}^{N}%
q_{j}x^{q_{j}-1}\right\rangle =\frac{\phi^{\prime}\left(  x\right)  }{\bar{q}%
}.\label{60}%
\end{equation} 
Note that $\bar{q}=\phi'(1)$. 

\begin{figure}[ptb]
\par
\begin{center}
\scalebox{0.33}{\includegraphics{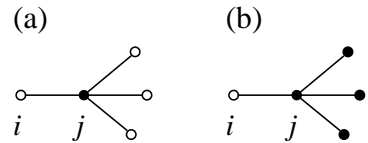}}
\end{center}
\caption{ The first (a) and the second (b) components of edge $(ij)$. Filled
vertices belong to the components. The $0$-th component is empty. }%
\label{f3}%
\end{figure}

Let the $n$-th component of the ordered pair $\left(  j,i\right)  $,
$C_{n,ji}$ be the following set of vertices. For any (ordered) pair of
vertices $\left(  j,i\right)  $, 
$C_{1,ji}$ is vertex $j$ 
if vertices
are connected, else $C_{1,ji}=\varnothing$. For $n>1$, $C_{n,ij}$ is defined
recursively as follows. If $g_{ji}=0$, all $C_{n,ji}=\varnothing$. Otherwise,
in $C_{2,ij}$ there are also $q_{j}-1$ other vertices, connected to the vertex
$j$, the third component $C_{3,ij}$ contains also all other vertices,
connected with ones of the second component, and so on, see Fig.~\ref{f3}.

In the thermodynamic limit ($N\rightarrow\infty$) almost every finite $n$-th
component of uncorrelated random graph is a tree. Then for the sizes (numbers
of vertices) of the components, $M_{n,ij}=\left\vert C_{n,ij}\right\vert $ we
have (assuming $g_{ji}=1$):
\begin{equation}
M_{n,ji}=1+\sum_{k\neq i}M_{n-1,kj},\label{80}%
\end{equation}
with the initial condition $M_{1,ji}=1$. Due to the absence of correlations in
the configuration model, $M_{n,kj}$ and $M_{n,lj}$, $k\neq l$, are independent
equally distributed random variables. We define the distribution function of
the $n$-th component of an edge as
\begin{align}
p_{n}(M) &  =\frac{1}{2L}\sum_{j,i=1}^{N}\langle g_{ji}\delta(M_{n,ij}%
-M)\rangle\nonumber\\[5pt]
&  =\frac{N(N-1)}{2L}\langle g_{ji}\delta(M_{n,ji}-M)\rangle\nonumber\\[5pt]
&  =\frac{1}{\langle g_{ji}\rangle}\langle g_{ji}\delta(M_{n,ji}%
-M)\rangle.\label{90}%
\end{align}
It is more convenient to use the Z-transformation of this distribution:%
\begin{equation}
\psi_{n}\left(  x\right)  =\frac{1}{2L}\left\langle \sum_{i,j=1}^{N}%
g_{ij}x^{M_{n,ij}}\right\rangle =\frac{1}{\left\langle g_{ij}\right\rangle
}\left\langle g_{ij}x^{M_{n,ij}}\right\rangle .\label{100}%
\end{equation}
Substituting Eq.~(\ref{80}) into Eq.~(\ref{100}) and using Eq.~(\ref{60})
gives
\begin{align}
&  \!\!\!\!\psi_{n}\left(  x\right)  =\frac
{x}{2L}\left\langle \sum_{j,i=1}^{N}g_{ji}\prod_{k\neq i,}\frac{1}%
{\left\langle g_{kj}\right\rangle }\left\langle g_{kj}x^{M_{n-1,kj}%
}\right\rangle \right\rangle \nonumber\\[5pt]
&  \!\!\!\!\!=\frac{x}{2L}\left\langle \sum
_{j=1}^{N}q_{j}\left[  \psi_{n-1}\left(  x\right)  \right]  ^{q_{j}%
-1}\right\rangle =x
\widetilde{\phi}
\left[  \psi_{n-1}\left(
x\right)  \right]  .\label{120}%
\end{align}
Let $\mathcal{C}_{n,i}$ be the $n$-th component of vertex $v_{i}$. This
component includes all vertices at distance $n$ or closer from vertex $v_{i}$.
(The $0$-th component of a vertex is empty.) Due to the absence of loops
(tree-like structure) we have the following relation for the size of $n$-th
component of vertex $v_{i}$, $\mathcal{M}_{n,i}=\left\vert \mathcal{C}%
_{n,i}\right\vert $,
\begin{equation}
\mathcal{M}_{n,i}=1+\sum_{j}M_{n-1,ij}.\label{140}%
\end{equation}
So the $n$-th component size distribution
\begin{equation}
P_{n}\left(  \mathcal{M}\right)  =\frac{1}{N}\left\langle \sum_{i=1}^{N}%
\delta\left(  \mathcal{M}_{n,i}-\mathcal{M}\right)  \right\rangle \label{150}%
\end{equation}
is expressed in Z-representation as
\begin{equation}
\Psi_{n}\left(  x\right)  =\frac{1}{N}\left\langle \sum_{i=1}^{N}%
x^{\mathcal{M}_{n,i}}\right\rangle =x\phi\left[  \psi_{n-1}\left(  x\right)
\right]  .\label{160}%
\end{equation}

The average sizes of subsequent $n$-th components are related through the
following equations:%
\begin{align}
\overline{\mathcal{M}}_{n} &  =\Psi_{n}^{\prime}\left(  1\right)  =1+\bar
{q}\,\overline{M}_{n-1},\label{170}\\[5pt]
\overline{M}_{n} &  =1+\zeta\overline{M}_{n-1},\label{180}%
\end{align}
where%
\begin{equation}
\zeta\equiv
\widetilde{\phi}\,' 
(1)=\frac{\phi^{\prime\prime}(1)}{\bar{q}}%
=\frac{1}{\bar{q}}\sum_{q=0}^{\infty}q(q{-}1)\Pi\left(  q\right)
=\frac{\langle q^{2} \rangle}{\bar{q}}-1,
\label{190}%
\end{equation}
which is the mean branching. 
If $\zeta<1$, both $M_{n}$ and $\mathcal{M}_{n}$ have finite limits as
$n\rightarrow\infty$. That is, the network has no giant connected component.
If $\zeta>1$, a giant connected component exists.

Assuming $\psi_{n}=\psi_{n-1}\equiv\psi$ in Eq.~(\ref{120}), we obtain an
equation for the distribution function of the sizes of edge's connected
components,
\begin{equation}
\psi(x)  =x
\widetilde{\phi}
[\psi(x)]  ,\label{200}%
\end{equation}
which implicitly defines $\psi(x)$. If $\zeta>1$, this equation
has two solutions at $x=1$. One is $\psi(1)=1$, the other is some
$\psi(1)\equiv t
<1$,
\begin{equation}
t
=
\widetilde{\phi}
(t
)  .\label{210}%
\end{equation}
For any value of $\zeta$, $\psi_{n}\left(  1\right)  =1$. On the other hand,
if $\zeta>1$, $\lim_{x\rightarrow1-0}\lim_{n\rightarrow\infty}\psi_{n}\left(
x\right)  =t
<1$. This is the probability that the connected component of a
randomly chosen edge is finite. Then the probability that randomly chosen
vertex belongs to a finite connected component of the graph is
\begin{equation}
\sum_{q=0}^{\infty}\Pi\left(  q\right)  t
^{q}=\phi\left(  t
\right)
.\label{215}%
\end{equation}
Therefore the number of vertices in the giant connected component in the
thermodynamical limit is
\begin{equation}
M_{\infty}=N\left[  1-\phi\left(  t
\right)  \right]  
.\label{220}%
\end{equation}

One may find an intervertex distance distribution from the mean sizes of the
$n$-th components of a vertex, see Ref.~\cite{nsw01,ch03}. The diameter
$\overline{\ell}$ of the giant connected component, i.e., the distance between
two randomly chosen vertices, is obtained from the relation $\zeta\sim
M_{\infty}\sim N$. So, if the second moment of the degree distribution is
finite,
\begin{equation}
\overline{\ell}\cong\frac{\ln aN}{\ln\zeta},\label{230}%
\end{equation}
where $a$ is some number of the order of $1$. For more straightforward
calculations, see Refs.~\cite{dms03,ffh04}.


\begin{thebibliography}{99}
\bibitem {remark1}One can easily understood the vertex equidistance in typical
infinite small worlds in the following way. The mean number of $\ell$-th
nearest neighbors of a vertex in small worlds usually grows with $\ell$ as
$\zeta$, where $\zeta$ is the mean branching coefficient, and $\ell
\lesssim\overline{\ell}(N)$. This leads to the following estimate for the
intervertex distance distribution: $\mathcal{P}(\ell,N)\propto\zeta
\theta(\overline{\ell}(N)-\ell)$. In addition, one must take into account
normalization, which readily gives $\mathcal{P}(\ell,N)\rightarrow\delta
(\ell-\overline{\ell}(N))$ as $N\rightarrow\infty$.
\end{thebibliography}
\end{document}